\documentclass[twocolumn,tighten]{aastex63}

\usepackage{graphicx}
\usepackage{ulem}
\usepackage{amsmath}
\usepackage{wrapfig}

\usepackage{tikz}
\usepackage{ulem}

\definecolor{tiffany}{RGB}{79, 166, 158}

\newcommand\lsim{\mathrel{\rlap{\lower4pt\hbox{\hskip1pt$\sim$}}
        \raise1pt\hbox{$<$}}}
\newcommand\gsim{\mathrel{\rlap{\lower4pt\hbox{\hskip1pt$\sim$}}
        \raise1pt\hbox{$>$}}}

\begin{document}
\shorttitle{Formation of massive black holes}
\shortauthors{Gonz\'{a}lez et al.}

\title{Intermediate-mass Black Holes from High Massive-star Binary Fractions in Young Star Clusters }

\correspondingauthor{Elena Gonz\'{a}lez}
\email{elenagonzalez@ufl.edu}

\author{Elena Gonz\'{a}lez}
\affil{Department of Astronomy, University of Florida, Gainesville, FL, 32611, USA}
\affil{Center for Interdisciplinary Exploration \& Research in Astrophysics (CIERA) and Department of Physics \& Astronomy, Northwestern University, Evanston, IL 60208, USA}

\author[0000-0002-4086-3180]{Kyle Kremer}
\affil{Center for Interdisciplinary Exploration \& Research in Astrophysics (CIERA) and Department of Physics \& Astronomy, Northwestern University, Evanston, IL 60208, USA}
\affiliation{TAPIR, California Institute of Technology, Pasadena, CA 91125, USA}
\affiliation{The Observatories of the Carnegie Institution for Science, Pasadena, CA 91101, USA}

\author[0000-0002-3680-2684]{Sourav Chatterjee}
\affil{Tata Institute of Fundamental Research, Homi Bhabha Road, Navy Nagar, Colaba, Mumbai 400005, India}

\author[0000-0002-7330-027X]{Giacomo Fragione}
\affil{Center for Interdisciplinary Exploration \& Research in Astrophysics (CIERA) and Department of Physics \& Astronomy, Northwestern University, Evanston, IL 60208, USA}

\author[0000-0003-4175-8881]{Carl L. Rodriguez}
\affil{McWilliams Center for Cosmology, Department of Physics, Carnegie Mellon University, Pittsburgh, PA 15213, USA}

\author[0000-0002-9660-9085]{Newlin C. Weatherford}
\affil{Center for Interdisciplinary Exploration \& Research in Astrophysics (CIERA) and Department of Physics \& Astronomy, Northwestern University, Evanston, IL 60208, USA}

\author[0000-0001-9582-881X]{Claire S. Ye}
\affil{Center for Interdisciplinary Exploration \& Research in Astrophysics (CIERA) and Department of Physics \& Astronomy, Northwestern University, Evanston, IL 60208, USA}

\author[0000-0002-7132-418X]{Frederic A. Rasio}
\affil{Center for Interdisciplinary Exploration \& Research in Astrophysics (CIERA) and Department of Physics \& Astronomy, Northwestern University, Evanston, IL 60208, USA}

\begin{abstract}

Black holes formed in dense star clusters, where dynamical interactions are frequent, may have fundamentally different properties than those formed through isolated stellar evolution. Theoretical models for single star evolution predict a gap in the black hole mass spectrum from roughly $40-120\,M_{\odot}$ caused by (pulsational) pair-instability supernovae. Motivated by the recent LIGO/Virgo event GW190521, we investigate whether black holes with masses within or in excess of this ``upper-mass gap'' can be formed dynamically in young star clusters through strong interactions of massive stars in binaries. We perform a set of $N$-body simulations using the \texttt{CMC} cluster-dynamics code to study the effects of the high-mass binary fraction on the formation and collision histories of the most massive stars and their remnants. We find that typical young star clusters with low metallicities and high binary fractions in massive stars can form several black holes in the upper-mass gap and often form at least one intermediate-mass black hole. These results provide strong evidence that dynamical interactions in young star clusters naturally lead to the formation of more massive black hole remnants.

\vspace{1cm}
\end{abstract}

\section{Introduction}
\label{sec:intro}

Since the first binary black hole (BBH) detection by LIGO/Virgo in 2015 \citep{Abbott2016}, the field of gravitational wave astrophysics has taken off. The recent release of the second Gravitational Wave Transient Catalog by LIGO/Virgo \citep{LIGO2020_O3} has increased significantly the number of known gravitational wave events, enabling new and powerful constraints on cosmology, fundamental physics, and various aspects of stellar astrophysics \citep[e.g.,][]{GWTC2_gr2020,GWTC2_grb2020}. In particular,
the ability to describe the population properties of the compact objects involved in the mergers has helped both confirm models for massive star evolution and test the limits of our theories \citep[e.g.,][]{LIGO2020_O3populations}.

Key evolutionary stages of massive stars expected to leave a strong imprint on the shape of the BH mass spectrum are pair-instability supernovae (PISNe) and pulsational pair-instability supernovae (PPISNe). These occur when evolved stars with core masses between roughly $45$ and $135\,M_{\odot}$ experience at the onset of carbon burning a decrease in radiation pressure and core contraction, associated with electron-positron pair production \citep[e.g.,][]{barkat}. Consequently, depending on the final stellar core mass, a series of runaway thermonuclear explosions can either significantly enhance mass loss (PPISN) or completely destroy the star, leaving behind no remnant (PISN) \citep{Fowler1964, Bond1984,Ober1983,HegerWoosley2002,Woosley2007,Belczynski2016b,Woosley2017,Woosley2019}. Recent calculations show that PISNe and PPISNe can prevent the formation of BHs with masses in the range $\approx40-120\,M_{\odot}$, with the exact boundaries remaining somewhat uncertain \citep[e.g.,][]{Woosley2017, Spera17, limongi2018, Takahashi2018, Stevenson2019, Marchant2019, Farmer2019, Mapelli2020, Renzo2020, Bel2020_pisne, Costa2020}.

GW190521, a BBH merger with component masses of about $66 M_{\odot} $ and $85 M_{\odot} $
\citep{Abbott}, revealed, for the first time, evidence of BHs with masses in the so-called ``upper mass gap'' expected from pair instabilities. In total, the first half of LIGO/Virgo's third observing run revealed 8~BBH mergers with at least one component with mass in excess of $45\,M_{\odot}$, 5~of which have at least one component in excess of $60\,M_{\odot}$. The direct detections of these massive BHs spark a number of questions concerning the stellar BH mass spectrum and prompts a detailed examination of our understanding of stellar evolution models.

A number of recent analyses have examined ways that BHs with masses residing in the upper-mass gap may form. Possibilities include hierarchical mergers of lower-mass BHs \citep[e.g.,][]{MillerHamilton2002,McKernan2012,Rodriguez2018b,Rodriguez2019,AntoniniGieles2019,GerosaBerti2019,FragioneLoeb2020,FragioneSilk2020}, primordial BHs formed through collapse of gravitational instabilities in the early universe \citep[e.g.,][]{LoebRasio1994,Carr2016}, Population III stars \citep[e.g.,][]{Madau_2001,BrommLarson2004}, growth through gas accretion in star forming environments \citep[e.g.,][]{Roupas2019}, and stellar mergers in dense star clusters \citep[e.g.,][]{DiCarlo19,Rizzuto,Kremer20,Banerjee_2020,Banerjee_2020_b}. In the latter case, a merger between a main-sequence star and an evolved star leads to the formation of a helium core that could survive the supernova explosion and result in a BH more massive than could ever be formed through single-star evolution \citep{Spera19}. This is closely related to the collisional runaway process that has long been associated with the formation of intermediate-mass black holes (IMBHs), with masses in the range $\approx10^2-10^5\,M_{\odot}$  \citep[e.g.,][]{Zwart,Gurkan,Giersz2015,Mapelli2016}. The existence of IMBHs in dense star clusters has been debated for decades, with possible evidence coming in the form of X-ray/radio sources and dynamical measurements \citep[for a recent review, see][]{Greene2020}. The roughly $150\,M_{\odot}$ BH observed as the merger product of GW190521 provides the first \textit{direct} evidence of IMBH formation. If indeed GW190521 was dynamically formed in a star cluster, it would be the strongest evidence yet that IMBHs, and massive BHs more broadly, can form in dense stellar environments.

In the dynamical evolution of young star clusters, one of the most crucial parameters pertaining to massive stars is the primordial binary fraction. For stars in the Galactic field, observations suggest that nearly $100\%$ of O- and B-type stars reside in binaries at birth \citep[e.g.,][]{Sana2012,MoeDiStefano2017}. The binary fraction (both primordial and at late times) in stellar clusters is less well constrained. Many old globular clusters are observed to have low binary fractions ($\lesssim10\%$) at present, even though their primordial binary fractions at birth may have been higher  \citep[e.g.,][]{Ivanova_2005,Milone2012}.

On the other hand, young massive clusters (YMCs) in the local universe have measured binary fractions comparable to those seen in the field \citep[e.g.,][]{Sana2009}. Progenitors of today's GCs are widely thought to have had properties similar to present-day YMCs, but with much lower metallicities \citep[e.g.,][]{Chatterjee_2010,Chatterjee2013}. Although a direct connection between YMCs and GCs remains elusive due to a lack of observed intermediate clusters \citep[e.g.,][]{PortegiesZwart2010}, the high observed binary fractions in YMCs suggest that a high binary fraction for massive stars may be present in all star clusters at birth.

The importance of stellar binaries for star cluster dynamics has been understood for decades, with binaries expected to serve as an important dynamical energy source and to slow down gravothermal contraction \citep[e.g.,][]{HeggieHut2003,Chatterjee2013,Chatterjee_2010}. Binaries also play a significant role in producing high rates of both stellar collisions \citep[e.g.,][]{Fregeau_2007} and BH mergers \citep[e.g.,][]{Chatterjee2017}.

In this study, we explore the effect of high-mass binary fraction upon the short-term evolution of dense star clusters with a specific focus on stellar collisions and the formation of massive BHs that may lie within or above the pair-instability mass gap.

This Letter is organized as follows. We describe our methods for modeling clusters in Section \hyperref[sec:method]{2}.  In Section \hyperref[sec:method]{3} we present our results from numerical simulations, including the formation paths for the most massive BHs. We also study the effects of the primordial high-mass binary fraction and the initial virial radius of clusters on BH formation. Finally, we discuss our results and conclude in Section \hyperref[sec:method]{4}.

\vspace{0.5cm}
\section{Models of Cluster Evolution}
\label{sec:method}

We perform numerical simulations using \texttt{CMC} (for \texttt{Cluster Monte Carlo}), a Hénon-type Monte Carlo code that models the evolution of stellar clusters \citep{Pattabiraman2013,Kremer20CMC}. This code incorporates prescriptions for various physical processes including two-body relaxation \citep{Joshi2000}, stellar/binary evolution using the population synthesis code \texttt{COSMIC} \citep{Breivik19}, direct integration of small-$N$ strong encounters using \texttt{Fewbody} \citep{Fregeau_2007}, and stellar collisions \citep{Fregeau_2007}. See Section 2.1 of \citet{Kremer20CMC} for a summary of \texttt{CMC} and Section 2 of \cite{Kremer20} for a detailed discussion of our treatment of collision cross sections, properties and evolution of collision products, and compact object formation for this set of simulations. A standard $\alpha\gamma$ model is adopted for common envelope (CE) evolution \citep{Hurley_2002}. The CE efficiency constant is $\alpha = 1.0$ and the binding-energy is set according to previous studies (see section 3.2 in \cite{Breivik19} for a recent overview of CE prescriptions in \texttt{COSMIC}).

The present study is based on the set of models listed in Table~\href{table: models}1. All models consist of $8\times10^5$ objects, corresponding to an initial total cluster mass of $4.7\times 10^5\,M_{\odot}$. The metallicity is set to $0.002$ ($0.1\,Z_{\odot}$)
and the initial conditions are King models with concentration parameter $W_0=5$. Stellar masses (primary masses for binaries) are sampled from a \citet{Kroupa2001} initial mass function in the range $0.08-150\,M_{\odot}$. To increase the robustness of our results, we run multiple realizations of each set of initial parameters with different random seeds.

We vary in our models the initial virial radius, $r_v$, and the high-mass binary fraction, $f_{\rm{b,high}}$, defined as the fraction of primaries with mass above $15\,M_{\odot}$ that have a companion at the time of cluster formation. The simulations performed include values for $r_v$ of $1,1.2$ and $1.5$ pc and  $f_{\rm{b,high}}$ of $0\%$ and $100\%$. For all models, the low mass ($<15\,M_{\odot}$) binary fraction is fixed at $5\%$. For low-mass binaries, primary masses are drawn randomly from our IMF, secondary masses are drawn assuming a flat mass ratio distribution in the range [0.1,1], and initial orbital periods are drawn from a log-uniform distribution $dn / d \log P \propto P$. For the secondaries of the massive stars ($> 15\,M_{\odot}$), a flat mass ratio distribution in the range [0.6,1] is assumed and initial orbital periods are drawn from the distribution
$dn / d \log P \propto P^{-0.55}$ \citep[e.g.,][]{Sana2012}.

For all binaries, the initial orbital periods are drawn from near contact to the hard-soft boundary and eccentricities are assumed to be thermal. The simulations are limited to 30~Myr as we focus on massive star evolution and BH formation.

In this paper we define ``pair-instability gap'' (or ``upper-mass gap'') BHs to be those with masses in the range $40.5-120\,M_{\odot}$, as determined by our assumed prescriptions for pair-instability physics \citep[for details, see][]{Belczynski2016b,Kremer20}. Although the pair-instability gap range considered is roughly consistent with that inferred from the latest LIGO/Virgo observations \citep{LIGO2020_O3populations}, we note that the exact boundaries of this gap are uncertain and depend upon various assumptions regarding massive star physics \citep[e.g.,][]{Woosley2017, Spera17, limongi2018, Takahashi2018, Stevenson2019, Marchant2019, Farmer2019, Mapelli2020, Renzo2020, Bel2020_pisne, Costa2020}. However, we stress that changes to the assumed boundaries of the pair-instability gap are unlikely to affect our results. Here we use the term ``IMBH'' to refer specifically to BHs with $M>120\,M_{\odot}$, beyond our assumed upper boundary for the pair-instability gap.
We use ``massive BH'' as a general term to refer to any BH with mass greater than $40.5\,M_{\odot}$.

\vspace{0.5cm}

\startlongtable
\begin{deluxetable*}{c|c|c|c|c|c|c|c|c|c|c}
\tabletypesize{\scriptsize}
\tablewidth{0pt}
\tablecaption{List of cluster models \label{table:models}}
\tablehead{
	\colhead{Model} &
	\colhead{$r_v$} &
	\colhead{$f_{\rm{b,high}}$} &
	\colhead{$N_{\rm{BH}}$} &
	\colhead{$N_{\rm{BH}}$}&
	\colhead{$N_{\rm{BH}}$} &
	\colhead{$N_{\rm{PI gap}}$} &
	\colhead{$N_{\rm{IMBH}}$} &
	\colhead{$M_{\rm{BH,max}}$} & 
	\multicolumn{2}{c}{Massive Star Mergers}\\
	\colhead{} &
	\colhead{(pc)} &
	\colhead{} &
	\colhead{(total)} &
	\colhead{(dyn coll.)} &
	\colhead{(bin coal.)} &
	\colhead{} &
	\colhead{} &
	\colhead{($M_{\odot}$)} &
	\colhead{(dyn coll.)} &
	\colhead{(bin coal.)} 
	}
\startdata
\texttt{1a} & 1 & 0 & 2259  & 240 & 0 & 0 & 0& 40.50 & 356 & 0  \\
\texttt{1b} & 1 & 0 & 2257  & 235 & 0 & 0 & 0& 40.50 & 390 & 0 \\
\texttt{1c} & 1 & 0 & 2254  & 232 & 0 & 1 & 0& 75.68 &  364 & 0\\
\texttt{1d} & 1 & 0 & 2259  & 241 & 0 & 0 & 0 & 40.50 &  390 & 0\\
\hline
\texttt{2a} & 1 & 1 & 2883  & 233 & 760 & 3 & 2 &598.28 &  656 &792 \\
\texttt{2b} & 1 & 1 & 2342  & 154 & 155 & 7 & 2 &230.29  & 409& 170 \\
\texttt{2c} & 1 & 1 & 2318  & 171 & 164 & 4 & 1 & 239.80  & 489 & 178\\
\texttt{2d} & 1 & 1 & 2185  & 158 & 160 & 7 & 0& 92.05  & 411 & 165\\
\texttt{2e} & 1 & 1 & 3223  & 240 & 155 & 7 & 2&  443.32  & 664 & 176\\
\texttt{2f} & 1 & 1 & 2588 & 175 & 167 & 5 & 2 &  279.24  & 474& 188 \\
\texttt{2g} & 1 & 1 & 2180 & 159 & 162 & 3 & 0 &  85.53  & 435& 172\\
\texttt{2h} & 1 & 1 & 2170  & 193 & 156& 3 & 0 & 82.45  & 722& 167 \\
\texttt{2i} & 1 & 1 & 2493 & 174 & 159 & 3 & 2 & 252.78  & 511& 177 \\
\texttt{2j} & 1 & 1 & 2169 & 177 & 158 & 2 & 0 &  90.97  & 476& 166 \\
\texttt{2k} & 1 & 1 & 3249 & 239 & 171& 11 & 2  &  468.84  & 607& 189  \\
\texttt{2l} & 1 & 1 & 2167 & 165 & 150 & 8 & 1 &  143.30  & 461& 162 \\
\texttt{2m} & 1 & 1 & 2214  & 139 & 146 & 4 & 1&  203.54  & 360& 165 \\
\texttt{2n} & 1 & 1 & 2152  & 152 & 155& 4 & 0 &  103.67  & 432& 167 \\
\texttt{2o} & 1 & 1 & 2185  & 146 & 149 & 2 & 0&  85.93  & 412 & 168 \\
\texttt{2p} & 1 & 1 & 2165  & 152 & 161 & 1 & 1&  154.85  & 443& 174 \\
\texttt{2q} & 1 & 1 & 2329 & 146 & 161 & 3 & 2  &  228.79 & 436 & 177 \\
\texttt{2r} & 1 & 1 & 2185  & 157 & 151 & 2 & 0 & 98.64 & 377 & 157 \\
\texttt{2s} & 1 & 1 & 2739 & 166 & 151& 0 & 1  & 303.11 & 516 & 174 \\
\texttt{2t} & 1 & 1 & 2192 & 138 & 160 & 7 & 0  &  99.20 & 345 & 164 \\
\texttt{2u} & 1 & 1 & 2175 & 154 & 150 & 4 & 0 &  96.43 & 439 & 172 \\
\texttt{2v} & 1 & 1 & 2172  & 121 & 154& 2 & 0 & 90.38 & 361 & 165 \\
\texttt{2w} & 1 & 1 & 2186  & 165 & 150 & 6 & 1 &  145.55 & 451 & 169 \\
\texttt{2x} & 1 & 1 & 2289  & 187 & 152 & 4 & 2 &  226.72 & 454 & 172 \\
\hline
\texttt{3a} & 1.2 & 1 & 2239 & 101 & 211 & 3 & 0 & 93.59 & 201 & 226 \\
\texttt{3b} & 1.2 & 1 & 2199 & 89 & 220 & 2 & 0 & 78.72 & 151 & 231 \\
\texttt{3c} & 1.2 & 1 & 2215  & 84 & 213 & 1 & 0&  49.41 & 161 & 226 \\
\texttt{3d} & 1.2 & 1 & 2221  & 94 & 210 & 2 & 0&  88.46 & 177 & 227 \\
\texttt{3e} & 1.2 & 1 & 2290  &  87 & 214 & 2 & 1&  217.44 & 185 & 228 \\
\texttt{3f} & 1.2 & 1 & 2183  & 84 & 216& 1 & 0 &  82.63 & 163 & 226 \\
\texttt{3g} & 1.2 & 1 & 2247 & 78 & 219 & 1 & 0  &  71.19 & 148 & 227 \\
\texttt{3h} & 1.2 & 1 & 2210 & 103 & 216 & 1 & 0  & 57.59 & 189 & 224 \\
\texttt{3i} & 1.2 & 1 & 2230  & 105 & 217 & 4 & 0& 99.08  & 187 & 227 \\
\hline
\texttt{4a} & 1.5 & 1 & 2335  & 45 & 621 & 2 & 1& 132.66  & 59& 625 \\
\texttt{4b} & 1.5 & 1 & 2321  & 40 & 623 & 0 & 0&  40.50  & 68& 631 \\
\texttt{4c} & 1.5 & 1 & 2357 & 38 & 621 & 0 & 0 & 40.50  & 54& 627 \\
\texttt{4d} & 1.5 & 1 & 2340  & 54 & 613& 0 & 0 & 40.50 & 75 & 624 \\
\texttt{4e} & 1.5 & 1 & 2356  & 43 & 619 & 1 & 0&  81.27  & 68& 626 \\
\enddata

\tablecomments{List of all cluster models included in this study. In column $2$ we indicate the initial virial radius in units of parsecs. Column $3$ lists the primordial high-mass binary fraction for the models. Column $4$ shows the total number of black holes formed. Columns $5-6$ show the number of black holes in the first two formation paths listed in Section \hyperref[sec:formation_channels]{3.1}. Columns $7-8$ indicate the number of black holes formed with masses in the pair-instability gap ($40.5 - 120\,M_{\odot}$) and number of IMBHs, respectively. Column $9$ lists the most massive black hole formed in solar masses. In columns $10$ and $11$ we show the total number of massive star collisions and massive binary coalescences where at least one mass component is massive ($M >15\,M_{\odot}$) and all components are either giants or main sequence stars. }
\end{deluxetable*}

\section{Results}
\label{sec:results}

\subsection{Formation channels}
\label{sec:formation_channels}

\begin{figure*}
\begin{center}
\includegraphics[width=1\linewidth]{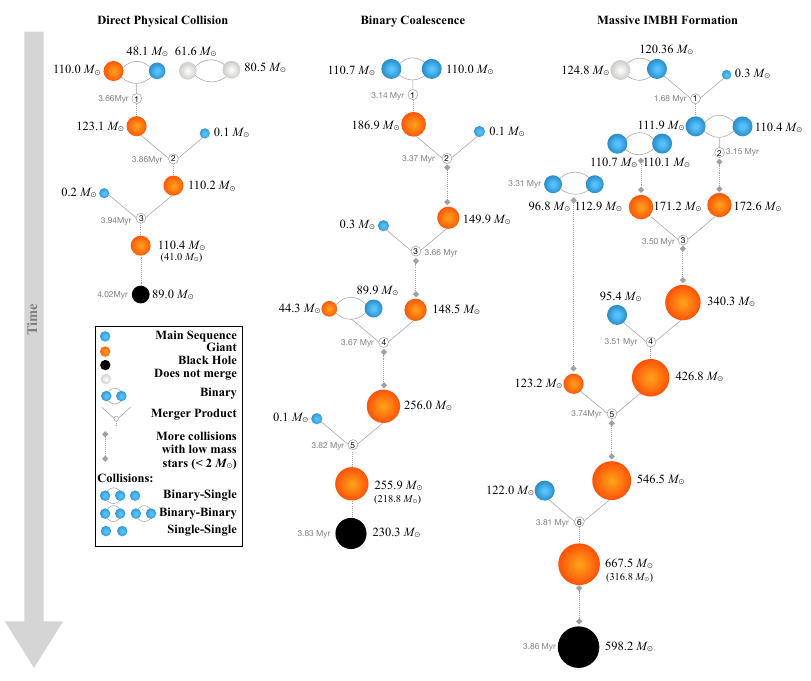}
\caption{\footnotesize \label{fig:tree}  The first two panels from the left illustrate the different formation paths described in section \hyperref[sec:formation_channels]{3.1}. In the direct physical collision scenario, the dominant process by which the principal object increases in mass is through physical collisions, as is shown by the initial binary-binary interaction that leads to the merger of two stars. In contrast, in the middle, for the formation path labeled binary coalescence, the main increase in mass comes from binary coalescence.The right-most column illustrates the collision history of the most massive IMBH formed in our simulations. We show the total mass of each object involved in the dynamical interactions. We show in parentheses the core mass just before BH formation.}

\end{center}
\end{figure*}

We distinguish formation channels based on the BH progenitor's primary method of growth. Illustrated in the two left-most panels of Figure \href{fig:tree}1 are characteristic examples of the two formation paths that we have identified in our simulations:

(1) \textit{Direct physical collisions:} The direct physical collision scenario occurs when most of the BH progenitor mass is accumulated via dynamical interactions (either single--single or binary-mediated) that lead to stellar collisions. In the example shown in the left-most panel of Figure \href{fig:tree}1, an initial binary-binary encounter leads to a collision and merger of one of the primordial binaries.
    
(2) \textit{Binary coalescence:} The second formation channel occurs when most of the mass growth comes from binary stellar evolution processes, in particular binary coalescence resulting from common envelope episodes. In this case, any subsequent dynamical collisions contribute less mass in total than the initial binary coalescence.
The middle panel in Figure \href{fig:tree}1 shows an example of a binary coalescence formation path.

We identify an additional formation path which is a subset of the ``direct physical collisions'' channel. In this scenario, which is illustrated in the right panel of Figure \href{fig:tree}1, the stars involved in the key dynamical collisions (e.g., collisions 3 and 5 in the right panel of Figure \ref{fig:tree}) are themselves products of primordial binary coalescences.

Although this ``hybrid'' channel is only observed in the formation history of three BHs in our models, it is notable because it highlights the interdependence of dynamics and binary evolution. Furthermore, it is the formation path of the most massive IMBH that we observe in this study (roughly 600 $M_{\odot}$).

Of all BHs formed in our models (fourth column of Table \ref{table:models}), $6.2$\% are formed through the direct physical collision path (column five) and $9.6$\% are labeled as binary coalescence (column six). The rest of the BHs are formed through standard isolated stellar evolution (i.e., experience no dynamical/binary evolution before BH formation).
For the massive BHs specifically, 95.9\% and  4.1\% fall into the direct collision and binary coalescence channels, respectively. For the massive BHs that are classified under the  direct collisions path, 100 \% feature at least one binary-mediated dynamical collision prior to formation, indicating the presence of primordial binaries plays a key role in massive BH formation. This is markedly different from the results of \citet{Kremer20} where, in the absence of primordial stellar binaries, mass growth through binary-evolution-mediated processes never occurred. The pronounced increase in the number of massive BHs in the presence of primordial binaries is the principal result of this study. 

\vskip 1.truein
\subsection{Primoridal high mass binary fraction }

To isolate the effect of $f_{\rm{b,high}}$, the top panel of  Figure \href{fig:bh_mass}2 compares the BH mass distributions for the models with $f_{\rm{b,high}}=0, 1$ and fixed $r_v=1\,$pc. The blue background marks the assumed boundaries for the pair-instability mass gap. The most notable difference between the two histograms is the presence of the extended tail of massive BHs ($M>40.5\,M_{\odot}$) in the model with $f_{\rm{b,high}}=1$ that is not present in the $f_{\rm{b,high}}=0$ case. As shown in Table \ref{table:models}, models with $f_{\rm{b,high}}=1$ (and $r_v=1\,$pc) produce, on average, $4$ BHs with masses in the range $40.5-120\,M_{\odot}$ (our assumed pair-instability gap) and $1$ BH with mass in excess of $120\,M_{\odot}$. In contrast, for the models with $f_{\rm{b,high}}=0$, we identify only $1$ pair-instability gap BH and zero BHs with masses larger than $120\,M_{\odot}$.\footnote{See \citet{Kremer20} for further information on the $f_{\rm{b,high}}=0$ models.}
It is clear from these results that the high-mass binary fraction makes a strong imprint on the BH mass distribution.

% \vskip 1.truein
\subsection{Virial radii }
In the bottom panel of Figure \href{fig:bh_mass}2, we plot the BH mass distributions for all of the $r_v$ models listed in Table \href{table:models}1 and fixed $f_{\rm{b,high}} = 1 $. It is clear that more compact clusters (smaller $r_v$) form a higher number of massive BHs, both within and above the mass gap. For example, in the models with $r_v=1\,$pc we see approximately 10 times more massive star collisions than in the models with $r_v=1.5\,$pc. This is to be expected since a denser environment will increase the rate of dynamical collisions of massive stars (see column $10$ in Table \href{table:models}1). The drastic change in mass of the heaviest BH formed between the models with virial radius $1$ pc and $1.5$ pc indicates that small changes in the density of a cluster can greatly influence the BH mass spectrum. As detailed in column $9$ in Table \href{table:models}1, the most massive BH formed in the model with $r_v=1\,$pc is approximately $600$ $M_{\odot}$ while the model with $r_v=1.5\,$pc has a maximum BH mass of $132$ $M_{\odot}$. Meanwhile, we observe a clear decrease in the rate of massive BH formation in models with $r_v>1.2\,$pc. 

\subsection{The role of high-mass binaries in BH formation}

Here we examine the different realizations of model \texttt{2} in Table \ref{table:models} ($r_v=1\,$pc and $f_{\rm{b,high}}=1$). As shown in Table \href{table:models}1, there is a difference in the number of BHs formed among the runs, with some producing as few as $2300$ BHs and other as many as  $3200$ BHs. Enhanced BH production occurs in the runs that form the most massive IMBHs, such as models \texttt{2a, 2e} and \texttt{2k}, which produce IMBHs ranging from roughly $400-600$  $M_{\odot}$.  The most massive IMBH in these models forms at $\sim$ $3$ Myr, while the increase in BH formation rate occurs between $7-15$ Myr.

This excess in BH number can be partially explained by an increase in binary interaction rates in these models. Specifically, model \texttt{2a} has a higher number of binary mergers involving stars with zero-age main sequence (ZAMS) masses between $10-25$ $M_{\odot}$ that would not have otherwise been  BH progenitors (at $0.1Z_{\odot}$, the minimum ZAMS mass that will form a BH is roughly $20\,M_{\odot}$). However, these merger products do evolve into BHs with masses roughly in the range of $M_{\rm{BH}} \approx 5-15\,M_{\odot}$. For models \texttt{2e} and \texttt{2k}, the higher number of BHs can be partially attributed to mass transfer in a binary increasing the mass of one of the components past the threshold of BH formation. Naturally, some of these differences are clearly due to stochastic fluctuations in the initial evolution of the model, leading to divergent paths in the dynamical evolution of the system. The main takeaway is that binary evolution processes, coupled with dynamics, can greatly influence BH growth. We leave a detailed study of the extent to which binary evolution processes affect the long-term cluster evolution and dynamics for a future study.

It is important to note that many (roughly $55\%$) of these ``excess'' BHs formed with masses less than  $15\,M_{\odot}$ are ejected promptly from their host cluster through natal kicks. Low-mass BHs are expected to form with less mass fallback and thus receive larger natal kicks \citep[e.g.,][]{Fryer2012}. Thus, this excess of BHs formed is unlikely to have a significant effect on the long-term cluster dynamics.
% \vskip 1.truein
\begin{figure}
\begin{center}
\includegraphics[width=0.9\linewidth]{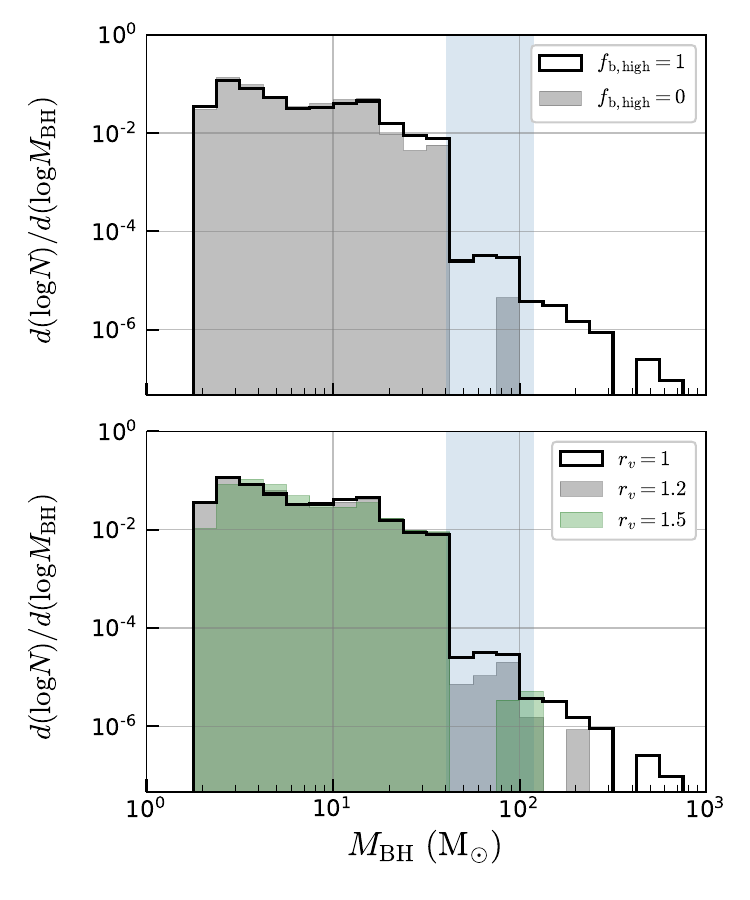}
\caption{\footnotesize \label{fig:bh_mass} Normalized black hole mass distribution for the models listed in Table \href{table: models}1. The top panel shows the black hole mass spectrum comparison between the two models with high mass binary fraction 0 and 1 and a virial radius of $1\,$pc. The bottom panel shows the black hole spectrum for the different values of $r_v$ (we use the initial virial radius of the assumed King model to determine the initial cluster density) and a fixed high mass binary fraction of 1. The shaded blue region indicates the ``upper mass gap.''}

\end{center}
\end{figure}

\vspace{0.5cm}
\section{Discussion and conclusions}
\label{sec:conclusion}

Our results show that high binary fractions for massive stars lead to an increase in binary-mediated dynamical interactions, which in turn have an important effect on massive BH formation. We have demonstrated that increasing the high-mass binary fraction, consistent with observations, while keeping all other cluster parameters (e.g., cluster masses, virial radii, metallicity) fixed at the values used in \citet{Kremer20CMC}, may significantly change the early evolution of YMCs. In particular, high-mass binaries facilitate high rates of massive star collisions (occurring through both dynamical encounters and binary-evolution-driven mergers) that can lead to the formation of massive BHs, both within and above the pair-instability mass gap. With the exception of the high-mass binary fraction, the parameters for the models calculated in this paper are identical to those in \cite{Kremer_MW}, which were shown to lead to present-day properties matching well those of Milky Way globular clusters.
If indeed the majority of high-mass stars have binary companions at birth, massive BH formation may therefore be a common occurrence in the early phases of globular cluster evolution.

As shown in Figure \ref{fig:bh_mass}, the formation of massive BHs is sensitive to small changes in initial virial radius. This is in agreement with previous works on the topic of IMBH formation which have shown that the onset of a collisional runaway occurs abruptly under small changes to cluster parameters \citep[e.g.,][]{PortegiesZwart2004,Gurkan2004,Giersz2015,Mapelli2016}. Indeed, the \texttt{CMC} models presented in \citet{Kremer20} exhibited a similar sharp dependence on $r_v$. However, \citet{Kremer20}, which assumed zero primordial binaries, found that the onset of collisional runaways occurred at $r_v\lesssim0.8\,$pc, while here the transition occurs at $r_v\approx1\,$pc. Not surprisingly, the presence of primordial binaries increases the collision rate and therefore moves the threshold for collisional runaway to larger cluster sizes.  \citet{Kremer20CMC} showed that varying $r_v$ from about $0.5-4\,$pc can explain naturally the full range of globular cluster properties observed at present in the Milky Way. Thus, small changes to the minimum $r_v$ value that lead to collisional runaways may have deep repercussions on the theoretical predictions for the presence of IMBHs in globular clusters.

In this study, we have focused on the first $30\,$Myr of cluster evolution to explore the formation of massive BHs. After their formation, these BHs can acquire, retain, and lose close companion stars and compact objects through few-body interactions \citep[e.g.,][]{sigurdsson1993,Morscher2015,macleod2016}. If the mass of one of the BHs is high enough, a cusp of objects can efficiently grow, affecting the innermost regions of the host cluster \citep[e.g.,][]{baum2005,heggie2007,lutz2013}. Some of the closest companions could form hierarchically separated binaries with the massive BHs and persist until they are replaced in few-body encounters \citep{frag2019}. Eventually, close interactions with the massive BHs can lead to the tidal disruption of a star or the inspiral and merger of a compact object \citep[e.g.,][]{haster2016,frag2018a,frag2018b}. We leave the detailed study of the long-term dynamics of massive BHs to future work.

The recently released data from the first half of LIGO/Virgo's third observing run provides strong evidence for the formation of BHs with masses in the pair-instability gap \citep{LIGO2020_O3,LIGO2020_O3populations}. Thus, the results of this study may have important implications for GW astrophysics.  In lieu of integration of the cluster models presented here over the full cluster lifetime ($\gtrsim10\,$Gyr), we perform an order-of-magnitude estimate to determine the rate of mass-gap BBH mergers. Consider here only those models with $r_v=1\,$pc. If we assume for simplicity that every mass-gap BH in our $f_{\rm{b,high}}=1$ models goes on to undergo one\footnote{In principle, a mass-gap BH could undergo more than one BH merger, if the merger products are retained in the cluster. However, given the large recoil kicks associated with GW emission, this has low probability and can be ignored to first approximation.} (dynamically-formed) BBH merger within a Hubble time \cite[as discussed in][this is a reasonable assumption]{Kremer20}, we predict, on average, $4$ mass-gap mergers per cluster.

The most massive BHs in a cluster will be the first to mass-segregate to the center, dynamically form binaries, and merge \citep[e.g.,][]{Morscher2015,Kremer20CMC}. Thus, as a crude approximation, we can assume that these BHs merge promptly (with negligible delay time). We can then estimate the volumetric rate of mass-gap mergers at redshift $z$ as
\begin{equation}
    \label{eq:rate}
    \Gamma(z) \approx  \frac{N_{\rm{gap}}}{M_{\rm{cl}}} \,  \rho_{\rm{SF}}(z) \, f_{\rm{SF}}
\end{equation}
where $N_{\rm{gap}}$ is the average number of pair-instability gap BHs formed per cluster (from our models, we find $N_{\rm{gap}}\approx 4$; again, we assume these also go on to undergo one BBH merger), $M_{\rm{cl}}=4.7\times 10^5\,M_{\odot}$ (the initial cluster mass assumed for our models), and $\rho_{\rm{SF}}(z)$ is the cosmological density of the star formation rate at redshift $z$. 
At $z=1$, when metallacities of $0.1Z_{\odot}$ (as assumed for the models in this study) are relevant, $\rho_{\rm{SF}}$ has a value of roughly $0.1\,M_{\odot}\,\rm{yr}^{-1}\,\rm{Mpc}^{-3}$ \citep[e.g.,][]{HopkinsBeacom2006}.
The (highly uncertain) factor $f_{\rm{SF}}$ is the fraction of the star formation assumed to occur in star clusters that may yield massive BH mergers. Taking $f_{\rm{SF}}$ as a free parameter for now, we estimate a mass-gap merger rate of roughly $f_{\rm{SF}} \times 100\,\rm{Gpc}^{-3}\,\rm{yr}^{-1}$ at $z\approx 1$ from young massive clusters. In contrast, for our models with $f_{\rm{b,high}}=0$, we find on average $0.2$ mass-gap BHs per cluster, which translates to a mass-gap merger rate of roughly $f_{\rm{SF}} \times 0.5\,\rm{Gpc}^{-3}\,\rm{yr}^{-1}$ at $z\approx 1$.

Observations suggest that the majority of stars form in stellar clusters or associations \citep[e.g.,][]{LadaLada2003} covering a wide mass range. In this study, we have modeled young clusters more representative of the high-mass tail of the cluster mass function, which is expected to scale as $M^{-2}$ \citep{LadaLada2003}.
It is not clear \textit{a priori} that the results from our models extend to lower cluster masses $\lesssim10^5\,M_{\odot}$, which dominate the cluster mass function by number. However, recent work by \citet{DiCarlo19} has shown that massive BHs within or above the pair-instability gap may form through stellar collisions and merge with other BHs in young clusters with masses as low as $10^3\,M_{\odot}$. As a simple estimate, we can take $10^3\,M_{\odot}$ as the minimum cluster mass yielding massive BH formation and merger. Assuming $10^2\,M_{\odot}$ as the minimum value for a cluster/association mass function covering the entire SFR \citep{LadaLada2003}, we can then estimate $f_{\rm{SF}}$ as

\begin{equation}
    f_{\rm{SF}} \approx \frac{\int_{10^3}^{10^6} M^{-2}\, dM}{\int_{10^2}^{10^6} M^{-2} \,dM} \approx 0.1,
\end{equation}
suggesting a merger rate of massive BHs of roughly  $10\,\rm{Gpc}^{-3}\,\rm{yr}^{-1}$ at $z\approx 1$. 

While this crude order-of-magnitude estimate is encouraging, a more systematic analysis will be necessary to make detailed predictions and determine their theoretical uncertainties. However regardless of details, it is clear that the primordial high-mass binary fraction in star clusters could play a key role in the formation of massive BH mergers, easily detectable as GW sources by LIGO/Virgo or future GW detectors.

% \vskip 1.truein
\section{acknowledgements} 

We are grateful to Mario Spera for key insights during the development of this project and we thank him for a careful reading of the manuscript. We thank the anonymous referee for helpful comments. This work was supported by NSF Grants AST-1757792 and AST-1716762 at Northwestern University. KK is supported by an NSF Astronomy and Astrophysics Postdoctoral Fellowship under award AST-2001751. GF acknowledges support from a CIERA Fellowship at Northwestern University. Computations were made possible through the resources and staff contributions provided for the Quest high-performance computing facility at Northwestern University. SC acknowledges  support  of  the  Department  of  AtomicEnergy,  Government  of  India,  under  project  no.   12-R\&D-TFR-5.02-0200.

% \software{\texttt{CMC} \citep{Joshi2000,Joshi2001,Fregeau2003, Fregeau2007, Chatterjee2010,Chatterjee2013,Umbreit2012,Morscher2013,Rodriguez2018b, Kremer2020}, \texttt{Fewbody} \citep{Fregeau2004}, \texttt{COSMIC} \citep{Breivik2020}}

\bibliographystyle{aasjournal}
\bibliography{papers}

\end{document}